\documentclass[10pt, preprint2]{aastex}

\usepackage{graphicx}

\begin{document}

\title{Selection Effects in the SDSS Quasar Sample: The Filter Gap Footprint} 

\author{M.B. Bell\altaffilmark{1}, S.P. Comeau\altaffilmark{1}}

\altaffiltext{1}{Herzberg Institute of Astrophysics,
National Research Council of Canada, 100 Sussex Drive, Ottawa,
ON, Canada K1A 0R6}

\begin{abstract}

In the Sloan Digital Sky Survey (SDSS) quasars are targeted using colors and anything that can cause the identifying characteristics of the colors to disappear can create problems in the source selection process. Quasar spectra contain strong emission lines that can seriously affect the colors in photometric systems in which the transmission characteristics vary abruptly and significantly with redshift. When a strong line crosses a gap between two filter passbands the color effects induced by the line change abruptly, and there is also a dimming in apparent brightness compared to those redshifts where the strong line is inside a filter passband where the transmission is high. The strong emission lines in quasars, combined with the varying detectability introduced by the transmission pattern of the five filters, will result in a filter-gap footprint being imprinted on the N(z) distribution, with more quasars being missed when a strong line falls in a filter gap. It is shown here that a periodicity of $\Delta$z $\sim0.6$ is imprinted on the redshift-number distribution by this selection effect. Because this effect cannot be rigorously corrected for, astronomers need to be aware of it in any investigation that uses the SDSS N(z) distribution. Its presence also means that the SDSS quasar data cannot be used either to confirm or to rule out the $\Delta$z $\sim0.6$ redshift period reported previously in other, unrelated quasar data.

\end{abstract}

\keywords{galaxies: active - galaxies: distances and redshifts - galaxies: quasars: general}

\section{Introduction}


In the Sloan Digital Sky Survey the quasar candidates that eventually make up the number distribution, N(z), are selected for optical spectroscopy via their colors in $ugriz$ broadband photometry. Because quasar spectra contain strong emission lines, a significant change in color can occur when a strong emission line crosses a gap between two adjacent filter bands. This can cause the quasar selection completeness to change with redshift. Similarly, reductions in apparent brightness that occur when a strong line crosses an insensitive portion of the filter response (a gap) can also be significant, and will cause some of the faintest sources to fall below the detection limit and be missed. Since both of the above effects will affect the completeness of the quasar selection process, they must also influence the final quasar number distribution, N(z). None of the sources that are selectively missed due to the above reasons during the candidate selection process can be recovered in later stages of the data processing. This means that the filter gap footprint will be irrevocably imprinted on SDSS quasar samples during the initial targeting process. Since the quasar redshifts are determined spectroscopically at a later stage, which uses only the i-band magnitudes, this stage cannot remove, or even mildly affect, the filter-gap footprint that was imprinted on the N(z) distribution during the initial quasar targeting process. Although making adjustments to the targeting algorithm might appear to remove structure in the N(z) distribution, it cannot recover the missing sources. 

Regions of quasar color space that contain a high density of stars (z = 2.7 and z = 3.5) or galaxies are also avoided by the quasar selection algorithm, and many of those that have colors that fall inside these regions will be missed in the selection process. However, exactly how the color changes affected the data below z = 2.2 appears not to have been completely understood initially since it was claimed that the selection algorithm was already quite complete below z = 2.2 by data release DR3, and that few sources were missed in this range \citep[see their Fig 6]{ric06}.

Recently it was found, however, that peaks and valleys in the N(z) distribution of 46,000 SDSS quasar redshifts do exist, and even showed evidence for a period of $\Delta$z $\sim0.6$ \citep{mcd06}. It was claimed that this could be further evidence for an intrinsic redshift component \citep{mcd06}. For this 0.6 periodicity to be caused by a selection effect which just by chance was identical to the intrinsic redshift quantization predicted several years earlier \citep{bel02a,bel02b,bel02c,bel03}, seemed very remote.

After evidence for a redshift periodicity near 0.6 was reported in the SDSS data \citep{mcd06} it was argued by \citet{ric06} and \citet{sch07} that the peaks and valleys in the N(z) distribution can be explained by selection effects. 

These authors showed that when the selection effects they proposed were taken into account, a redshift distribution was obtained that no longer showed peaks and valleys. According to these authors their process accounted for emission-line effects, and used only objects selected in regions where spectroscopic observations were chosen with the final version of the quasar target selection algorithm. However, their correction process removed close to one-half of the sources, and no visible connection was made as to how the selection effects were tied to the peaks and valleys, or to how the $\Delta$z $\sim 0.6$ period might have been created.



Here we take a close look at how the spacing and widths of the five filters used in the SDSS survey might influence quasar selection using colors, and show that a redshift periodicity of $\Delta$z $\sim0.6$ that is directly related to the filter transmission function will be produced in the N(z) distribution obtained in the SDSS quasar sample.
 


\section{Transmission Characteristics of the SDSS Filters}

In the SDSS quasar target selection process five filters with similar widths centered at wavelengths between $\lambda$3000\AA\ and $\lambda$9000\AA\ are involved. These are referred to as $u^\prime$, $g^\prime$, $r^\prime$, $i^\prime$, and $z^\prime$ filters and their widths and locations are shown in Fig 1, which has been reproduced from Fig 9 of \citet{ric02}. In this figure there are four regions of lower sensitivity clearly visible at the intersections of the five bands, although the central two are significantly deeper. If there is a strong emission line present in the optical spectrum of quasars it will move across the filter bands from short to long wavelengths as quasars with higher and higher redshifts are observed \citep[see their Fig 2]{kac09}.

In order to determine which emission lines would likely produce the largest effect we used the composite quasar spectra reported previously by several groups \citep{fra91,zhe97,van01}. We limited our quasar sample to sources with z $>0.3$. Below z = 0.3 the quasar number distribution can be contaminated with non-quasars such as N-galaxies and BL Lacs. Because of the sharp decrease in source numbers above z $\sim2$, we also restricted the N(z) analysis to sources with z $< 2.4$.

In Table 1 we list the strongest emission lines that will cross at least one of the two deepest gaps in the above redshift range. Although Ly$\alpha$ and C IV$\lambda$1549\AA $\ $ are not part of this list, the former has been included simply because it is the intensity reference, and C IV$\lambda$1549\AA $\ $ has also been included for reasons that will become obvious below. The relative composite line strengths reported for these lines are given in columns 3, 4, and 5, and the mean value is listed in column 6. Because there are sometimes large differences in line intensities between these three investigations, we consider the mean Flux value to be the most reliable. 

In Fig 2, the paths followed by the ions listed in Table 1 are plotted versus redshift. Also indicated in Fig 2 by the horizontal lines are the locations of all four gaps that occur at the intersection of two adjacent filters. From this plot it is then easy to determine, for each emission line, the exact redshift at which each line will cross each filter gap. These redshifts are listed in Table 2 for those lines that cross at least one of the two deepest filter gaps. The strongest line inside the observing window, and that crosses at least one of the two deep gaps near 5450\AA $\ $ and 6800\AA, is Mg II$\lambda$2798\AA. It is also the only strong line that crosses all four filter gaps in the redshift range between z = 0.3 and 2.4.

The vertical gray lines in Fig 2 indicate the redshifts of 0.9, 1.4 and 2.0, where valleys were previously reported in the number distribution below z = 2.4 \citep[see Fig 5 of that paper]{mcd06}. It is obvious from the location of the gray lines in Fig 2, and from the redshifts listed in Table 2, that there is a valley in the number distribution located near each redshift where the strongest line (MgII $\lambda$2798\AA) crosses a filter gap. The best fits are for the valleys near z = 0.9 and 1.4, where MgII crosses the two deepest gaps. 

Also indicated in Fig 2 by the vertical dashed lines are the locations of the deep valleys near z = 2.7 and 3.5. These valleys in the N(z) distribution of quasars have been discussed and accounted for previously by a similar effect related to color confusion with stars \citep{ric02,ric06,mcd06}. It is apparent in Fig 2 that these valleys can also be explained by the two strongest lines (CIV and Ly$\alpha$) passing through the two deepest filter gaps. Although C IV appears to predict a valley closer to z = 2.5 this may be due to the fact that this valley is located near the steep drop-off in source number above z = 2.

\section{Evidence that Strong Emission Lines Can Seriously Affect Quasar Colors}
 
When colors are used to target quasars, any color changes that occur when redshift changes move a strong line from the peak of a filter passband to a valley between two passbands are likely to be important. The effect will be greatest for the strongest emission lines. To demonstrate that the color change can indeed be significant the plot in Fig 3 shows how the (u*-g*) and (g*-r*) colors change with redshift. In these curves the square-wave modulation is produced by the color-polarity change that occurs when MgII (the strongest line in this redshift range) moves from one filter bank to the next. These cross-over points are indicated by the strong, vertical lines, and the arrows indicate the redshifts where valleys were previously found in the N(z) distribution \citep{mcd06}.

 In Fig 4, where the square of the color difference is plotted versus redshift, it is demonstrated that when the line crosses a gap the difference between these two colors drops to near zero. The square is used here because we are only interested in the magnitude of the color differences. In Fig 4 the vertical bars again indicate the redshifts where MgII crosses the gaps between the filters, and the vertical arrows indicate, as in Fig 3, the locations where valleys were previously found in the N(z) distribution \citep{mcd06}. These results indicate that sources in these redshift ranges may be harder to identify as quasars when the large difference in colors seen when the emission line is in the middle of a filter passband, disappears. Alternatively, it may indicate that because this is a detection-limited sample, more sources are simply missed because they appear fainter when the strong emission line is in a gap causing more of them to fall below the detection limit. 





It is thus clear from Figs 3 and 4 that the filter transmission characteristics can have a large effect on the colors when strong emission lines are present in the spectrum, resulting in some sources being missed when a strong line crosses gaps between the filter passbands where the transmission is low and color differences disappear. Because of this effect an imprint related to the shape of the filter transmission function (a filter gap footprint) will be imprinted on the N(z) distribution during the quasar targeting process.



\subsection{Convolution of the Filter Gap Transmission Function With the Distribution of Strong Emission Lines}

In order to examine this effect from a different angle, in Fig 5 we have convolved the filter gap function shown in Fig 1 with the distribution of strong emission lines listed in Table 1. In this figure the convolution result is plotted versus redshift. Here, the convolution result is defined as the scalar product obtained by convolving the filter passband pattern with strong emission lines, where the lines are approximated by step functions of intensity equal to the line strength and width equal to the equivalent width. In Fig 5 the dashed lines have spacings of $\Delta$z = 0.6. There are several things to note from this result. First, there is a peak in the convolution function at redshifts predicted for a periodicity near 0.6, which confirms that the filter gap footprint can produce a periodicity in the N(z) distribution similar to that reported previously. Second, because the redshift range has been extended here to z = 4.6, the effects produced by L$_{\alpha}$ can also be determined. The peaks seen at z = 3.0 and 4.2 are produced when L$_{\alpha}$ is near the center of the g* and r* filters respectively, which again suggests that, in addition to the color confusion with stars, the filter transmission characteristics may also contribute to the valleys seen in the N(z) distribution near z = 2.7 and 3.5.

\section{Effects of the Filter Gap Footprint on the Accuracy of the Photometric Redshifts}

The photometric redshift, zp, is an approximate estimate of the redshift based on broadband photometric observables such as magnitudes and colors. Because of the obvious effect on the colors demonstrated above, it is likely that the confidence that can be placed on the photometric redshift will also vary with redshift.

 If the quasi-periodic structure in the N(z) distribution is color related, the same structure should be present in the N(zphot) distribution, where zphot is an estimate of the reliability of the photometric redshift as a function of redshift (the fraction of zp redshifts that are estimated to be correct). Recently the photometric redshifts of quasars have been discussed by \citet{ric09}. In Fig 6 we have plotted zphot versus redshift, for redshifts below z = 3.4 where the source density is high. This plot was taken from Fig 14 of \citet{ric09}. The dashed lines in Fig 6 again indicate redshifts with spacings of z = 0.6. It is clear that there is a peak at each of these locations.

\subsection{Effects of the Filter Gap Footprint on zConf}
 
An estimate of the reliability of each SDSS quasar redshift has been included in SDSS data releases DR5 and later. This number is referred to as zConf and, like zphot, is also related to the uncertainty introduced by the filter transmission function shown here in Fig 1. In Fig 7 the upper curve shows the mean zConf value for all sources in the DR6 catalogue \citep{ade08} plotted versus redshift. The zConf values can be accessed from the SDSS database server via a direct SQL query and individual values have been plotted versus redshift by \citet{har09}. The lower curve in Fig 7 is a plot of the filter transmission function in Fig 1 convolved with the MgII lines. The vertical bars at the top indicate where MgII crosses the gaps between the filters, indicating beyond any doubt that the filter gap footprint is very clearly present in the zConf data, and is very likely the source of the \em unknown selection effect \em discussed by \citet{har09}.

\section{Discussion}
 
In Fig 8 we show the Fourier transform of the plot in Fig 6, and it can be seen that there is a strong peak at $\Delta$z = 0.62. Fig 9 has been copied from \citet{mcd06} and shows a similar power peak at $\Delta$z = 0.62 that was obtained using the N(z) distribution as described in \citet{mcd06}.
We conclude from the good agreement obtained between the N(z) periodicity and the filter gap footprint structure that the periodicity is due to the filter gap footprint. A similar periodicity was reported by \citet{har09} for the zConf distribution. Although it has been claimed \citep{tan05} that there is no evidence for a periodicity in the number distribution of SDSS quasars, clearly this is not the case since we have seen above that the filter gap footprint, producing a period near $\Delta$z $\sim0.6$, is ubiquitous.

We conclude that attempts to detect a redshift quantization in the SDSS quasar sample with a period close to 0.6, that might originate from something other than the filter gap footprint, will not be possible in the presence of this strong selection effect. Thus, it will not be possible to use the raw SDSS quasar samples either to confirm, or to rule out, the redshift periodicity near 0.6 reported previously in completely unrelated quasar data \citep{bel02a,bel02b,bel02c,bel03}. Before this can be attempted it will be necessary for this selection effect to be rigorously accounted for. As noted above, attempts have been made by members of the SDSS group to correct for selection effects, and these appear to be able to remove all peaks and valleys in the number distribution \citep{ric06,sch07}. However, in their description of how corrections were made to the data to achieve this they do not give sufficient detail to allow the reader to reproduce their result. It appears to have been accomplished, at least in part, by repeated adjustments to the target selection algorithm. However, because several attempts were required it is also apparent that either, a), a rigorous correction is not possible or, b), a second component is present in the data with a different origin but with a similar periodicity. This approach does not appear to allow for the possibility, however slight, that an additional, unknown source of quantization might be present with the same period. In fact, it would seem that the only accurate way of correcting for this selection effect would be not to throw away one-half of the sources, but to find the missing sources by repeating the observations using a new set of filters whose passbands are centered at the locations of the present gaps. However, this is unlikely to happen, and it would certainly require a lot more evidence than currently exists before it would be worth pursuing.

\section{Conclusion}

We have demonstrated that during the SDSS quasar targeting process the filter transmission function imprints on the quasar N(z) distribution a pattern we refer to here as the filter gap footprint. This footprint is due to brightness and color changes related to the passage of strong emission lines (particularly Mg II$\lambda$2798\AA, below z = 2.2) across the five SDSS filters whose transmission function varies strongly with redshift. Although the redshifts are later measured spectroscopically, because this stage of the data acquisition process uses only the i-band data, it cannot significantly affect the filter gap footprint that was imprinted on the N(z) distribution during the initial quasar targeting stage. We show that this footprint introduces a periodicity near $\Delta$z $\sim0.6$ into the N(z) distribution. Because this effect cannot be rigorously corrected for, astronomers need to be aware of it in any investigation that uses the SDSS N(z) distribution. Its presence also means that the SDSS quasar data cannot be used either to confirm or to rule out the $\Delta$z $\sim0.6$ redshift period reported previously in other, unrelated quasar data.

\section{Acknowledgements}

We thank Don McDiarmid for many useful suggestions, and Gordon Richards of the SDSS Group for helpful information concerning the SDSS data acquisition system. 

Funding for the SDSS and SDSS-II has been provided by the Alfred P. Sloan Foundation, the Participating Institutions, the National Science Foundation, the U.S. Department of Energy, the National Aeronautics and Space Administration, the Japanese Monbukagakusho, the Max Planck Society, and the Higher Education Funding Council for England. The SDSS Web Site is http://www.sdss.org/.
The SDSS is managed by the Astrophysical Research Consortium for the Participating Institutions. The Participating Institutions are the American Museum of Natural History, Astrophysical Institute Potsdam, University of Basel, University of Cambridge, Case Western Reserve University, University of Chicago, Drexel University, Fermilab, the Institute for Advanced Study, the Japan Participation Group, Johns Hopkins University, the Joint Institute for Nuclear Astrophysics, the Kavli Institute for Particle Astrophysics and Cosmology, the Korean Scientist Group, the Chinese Academy of Sciences (LAMOST), Los Alamos National Laboratory, the Max-Planck-Institute for Astronomy (MPIA), the Max-Planck-Institute for Astrophysics (MPA), New Mexico State University, Ohio State University, University of Pittsburgh, University of Portsmouth, Princeton University, the United States Naval Observatory, and the University of Washington.

\clearpage

\begin{deluxetable}{cccccc}
\tabletypesize{\scriptsize}
\tablecaption{Relative Flux of Strong Emission Lines in Observing Window. \label{tbl-1}}
\tablewidth{0pt}
\tablehead{
\colhead{Emission Line} & \colhead{Rest $\lambda$} & \colhead{F(Vanden Berk)} & \colhead{F(Zheng)} & \colhead{F(Francis)} & \colhead{F mean}
}

\startdata

Ly$\alpha$ & $\lambda$1216\AA & 100 & 100 & 100 & 100 \\
C IV & $\lambda$1549\AA & 25.29 & 62 & 63 & 50.1 \\
Mg II & $\lambda$2798\AA & 14.7 & 25 & 34 & 24.5 \\

C III & $\lambda$1909\AA  & 15.9 & 11.5 & 29 & 18.1 \\
H$\beta$ & $\lambda$4861\AA & 8.6 & outside & 22 & 15.5 \\

\enddata 

\end{deluxetable}

\begin{deluxetable}{ccccc}
\tabletypesize{\scriptsize}
\tablecaption{Redshifts where strong emission lines fall at gaps between the filters. \label{tbl-2}}
\tablewidth{0pt}
\tablehead{
\colhead{Emission Line} & \colhead{Redshift ($u^\prime$-$g^\prime$)}  & \colhead{Redshift ($g^\prime$-$r^\prime$)} & \colhead{Redshift ($r^\prime$-$i^\prime$)} & \colhead{Redshift ($i^\prime$-$z^\prime$)} 
}

\startdata

Ly$\alpha \lambda$1216\AA & 2.2 & 3.48 & --- & --- \\
C IV $\lambda$1549\AA & 1.52 & 2.52 & 3.4 & --- \\

Mg II $\lambda$2798\AA &  0.39 & 0.95 & 1.43 & 1.96 \\
C III $\lambda$1909\AA  & 1.04 & 1.85 &  2.56  & 3.33 \\
H$\beta \lambda$4861\AA & --- & 0.1 & 0.4  & 0.7 \\
\enddata 

\end{deluxetable}

\clearpage

\begin{figure}
\hspace{-2.0cm}
\vspace{-2.5cm}
\epsscale{1.0}
\plotone{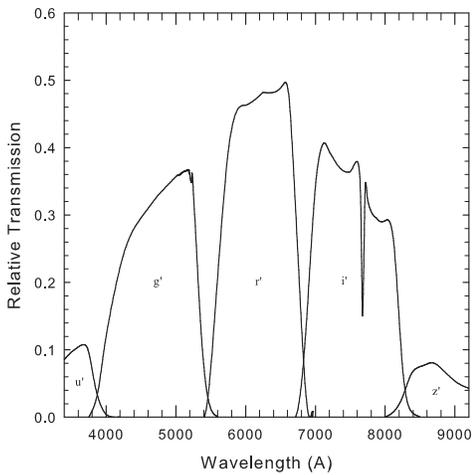}
\caption{{Plot showing the 5 SDSS filter passbands with gaps at their intersections. The two deepest gaps occur at 5450\AA\  and 6800\AA. \label{fig1}}}
\end{figure} 
 
\begin{figure}
\hspace{-1.0cm}
\vspace{-0.9cm}
\epsscale{0.9}
\plotone{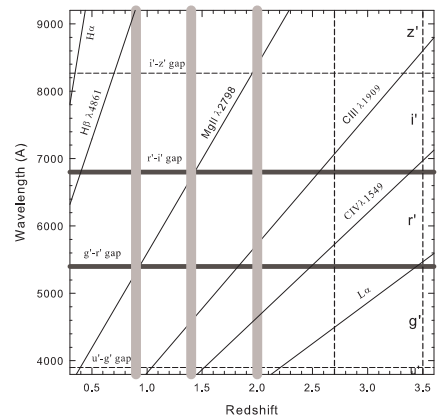}
\caption{{(solid lines)Paths followed by the strongest emission lines as they are redshifted through the SDSS filter passbands. The dark horizontal lines indicate the locations of the two strongest gaps in Fig 1, and the two dashed lines represent the two weaker gaps. The vertical gray lines indicate the location of valleys reported previously by \citet{mcd06}. The vertical lines at z = 2.7 and 3.5 correspond to the valleys discussed by \citet{ric02}. \label{fig2}}}
\end{figure}

\begin{figure}
\hspace{-1.0cm}
\vspace{-0.9cm}
\epsscale{0.9}
\plotone{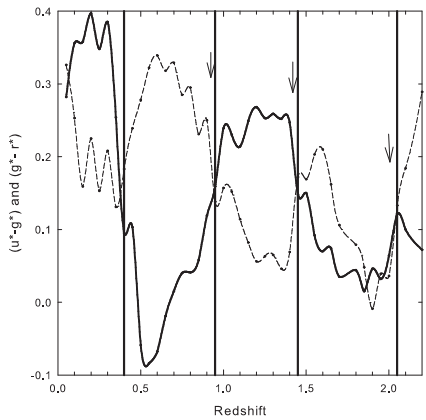}
\caption{{Median quasar colors (solid line for (u*-g*) and dashed line for (g*-r*)) plotted vs redshift from \citet{ric01}. The vertical solid lines indicate the four redshifts where the strongest emission line in this redshift range (MgII) crosses a filter gap. Arrows indicate where valleys were found in the N(z) distribution \citep{mcd06}.
\label{fig3}}}
\end{figure}

\begin{figure}
\hspace{-1.0cm}
\vspace{-0.9cm}
\epsscale{0.9}
\plotone{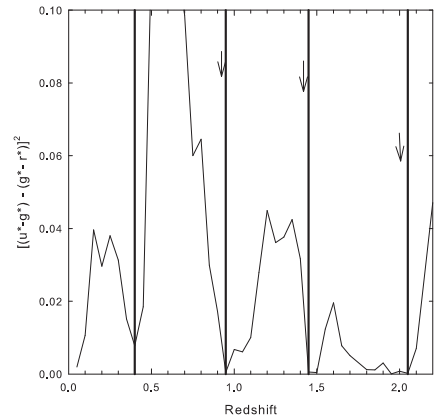}
\caption{{Plot showing where differences between (u$^{*}$-g$^{*}$) and (g$^{*}$-r$^{*}$) colors are small. The vertical lines indicate the redshift where MgII crosses the gaps between the filters. The arrows indicate the redshifts where valleys were found in the N(z) distribution. \label{fig4}}}
\end{figure}

\begin{figure}
\hspace{-1.0cm}
\vspace{-0.9cm}
\epsscale{0.9}
\plotone{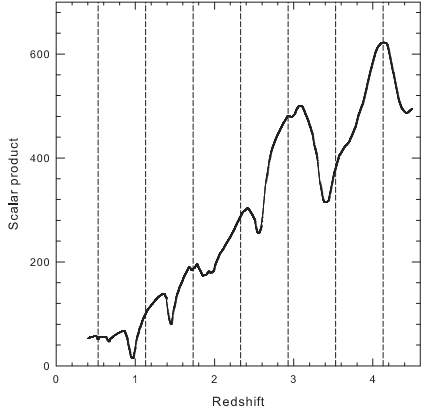}
\caption{{Here the scalar product is calculated as described in the text and plotted over a range of redshift from z = 0.3 to z = 4.5. The dashed lines show where peaks would be expected for a periodicity in z of 0.6. \label{fig5}}}
\end{figure}

\begin{figure}
\hspace{-1.0cm}
\vspace{-0.5cm}
\epsscale{0.8}
\plotone{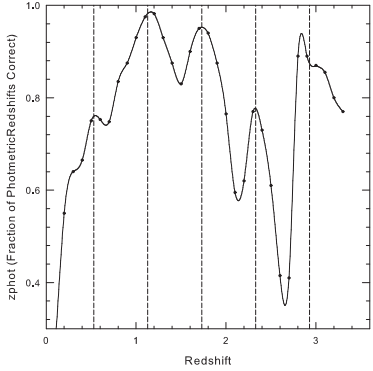}
\caption{{A plot showing the fraction of zphot redshifts that are estimated to be correct \citep{ric09} plotted versus redshift. The dashed lines show where peaks would be expected for a periodicity in z of 0.6. \label{fig6}}}
\end{figure}

\begin{figure}
\hspace{-1.0cm}
\vspace{-0.9cm}
\epsscale{0.9}
\plotone{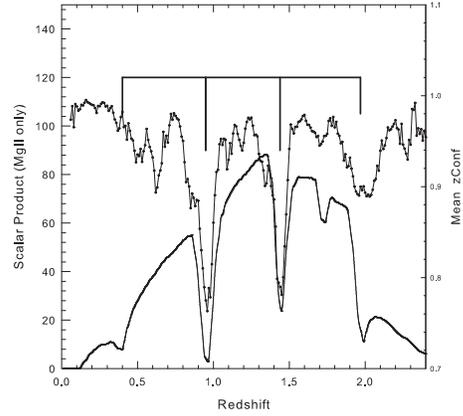}
\caption{{(upper curve)Mean zConf plotted vs z. (lower curve) Plot of the filter passbands convolved with the MgII $\lambda$2798 lines. The vertical bars indicate the redshift where MgII crosses the gaps between the filters. \label{fig7}}}
\end{figure}

\begin{figure}
\hspace{-1.0cm}
\vspace{-0.9cm}
\epsscale{0.9}
\plotone{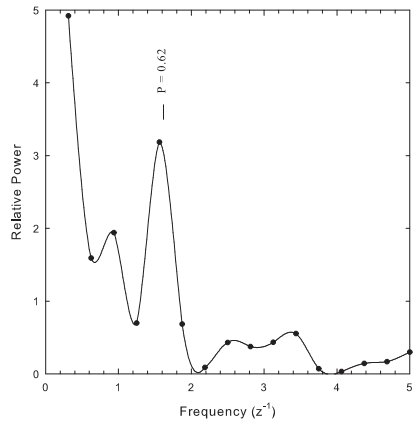}
\caption{{Fourier transform of the zphot plot in Fig 6 showing a strong power peak for a periodicity near $\Delta$z = 0.62. \label{fig8}}}
\end{figure}

\begin{figure}
\hspace{-1.0cm}
\vspace{-0.9cm}

\epsscale{0.9}
\plotone{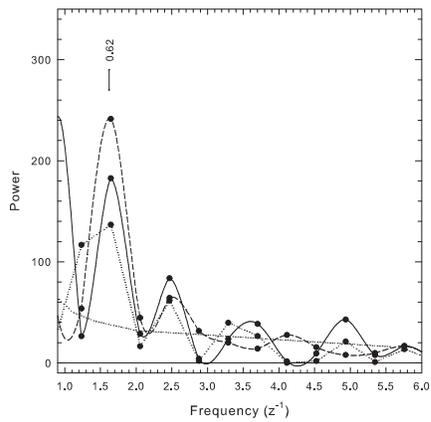}
\caption{{Fourier transform of the N(z) distribution showing a similar power peak for a periodicity near $\Delta$z = 0.6. See \citet{mcd06} for an explanation of the different curves. \label{fig9}}}
\end{figure}

\end{document}